# Putting a Price on Immobility: Food Deliveries and Pricing Approaches

Runyu Wang, Haotian Zhong*

School of Public Administration and Policy Renmin University of China

Emails: wangrunyu@ruc.edu.cn; hzhong@ruc.edu.cn

*Corresponding author

**Abstract**

Urban food delivery services have become an integral part of daily life, yet their mobility and environmental externalities remain poorly addressed by planners. Most studies neglect whether consumers pay enough to internalize the broader social costs of these services. This study quantifies the value of access to and use of food delivery services in Beijing, China, through two discrete choice experiments. The first measures willingness to accept compensation for giving up access, with a median value of ¥588 (≈$80). The second captures willingness to pay for reduced waiting time and improved reliability, showing valuations far exceeding typical delivery fees (e.g., ¥96.6/hour and ¥4.83/min at work). These results suggest a substantial consumer surplus and a clear underpricing problem. These findings highlight the need for urban planning to integrate digital service economies into pricing and mobility frameworks. We propose a quantity-based pricing model that targets delivery speed rather than order volume, addressing the primary source of externalities while maintaining net welfare gains. This approach offers a pragmatic, equity-conscious strategy to curb delivery-related congestion, emissions, and safety risks, especially in dense urban cores.

## 1. Introduction

Food delivery services have become an integral part of today's (im)mobility infrastructure (Dablanc et al., 2017; Xiang, 2024). However, such seemingly emancipatory services have led to negative externalities, including traffic accidents, congestion, pollution, and deteriorating labor conditions(Allen et al., 2021; Dong et al., 2021; Huang, 2022; Lezcano et al., 2023; Van Doorn, 2017). A key driver of these externalities is the relentless pursuit of speed, which is facilitated by the algorithms of commercial platforms that manage delivery workers' movements(Sun, 2019). To meet the strict delivery deadlines, measured in minutes, delivery workers often engage in risky behaviors such as running red lights, speeding, and riding against traffic. Because of these externalities, consumers do not bear the full social marginal costs of using food delivery services, especially in Chinese cities, where delivery fees are often reduced to near zero through platform subsidies and promotional waivers due to competitions among different platforms.

Despite the aforementioned negative externalities, urban food delivery services also improve economic welfare and generate utility. First, these platforms provide ordering services at no direct cost to users. They allow residents to meet daily needs or adapt to crises without the need for physical movement (Wang & He, 2021; Zhang et al., 2023; Y. Zhang et al., 2025). This type of free digital goods contribute a large amount of consumer surplus without positive prices (Brynjolfsson et al., 2019). Second, food delivery enables consumers to allocate more time to paid work or leisure activities, producing gains in welfare and utility. These welfare effects can be instrumental in designing pricing schemes to correct externalities. For instance, a potential Pareto improvement could be realized by raising delivery fees to a level below consumers' welfare gains, thereby enhancing overall efficiency without reducing individual well-being. Nevertheless, these welfare gains remain insufficiently quantified and warrant further investigation.

In this paper, we quantify the value of accessing and using food delivery services through two distinct choice experiments, single binary discrete choice (SBDC) experiments and multiple discrete choice (MDC) experiments. The SBDC experiment elicits willingness-to-accept (WTA) valuations, estimating a median monthly value of ¥586—the compensation an individual would require to forgo food delivery for one month. The MDC experiment measures willingness-to-pay

(WTP) for reduced waiting times and improved reliability. In the work scenario, the average WTP is ¥96.6 per hour saved and ¥4.83 per minute reduction in unreliability; in the home scenario, the figures are ¥99.6 per hour and ¥6.68 per minute, respectively. Finally, we calculate the aggregate social welfare change for our sample. The results indicate a total welfare gain from delivery time savings of approximately ¥60,504 per hour, or about ¥1,008 per minute. These estimates provide novel evidence that consumers place a substantial monetary value on urban food delivery services, which far exceed their direct costs. This difference generates a consumer surplus with implications for pricing strategies.

The rest of this paper is organized as follows. Section 2 reviews the literature on immobility, urban freight, and pricing in transportation. Section 3 introduces the empirical context, survey design, and analytic approach. Section 4 presents the results of the valuation experiments, preference heterogeneity, and welfare analysis. Section 5 concludes by discussing policy implications, limitations, and avenues for future research.

1. **Putting a Price on Immobility**

**2.1 Digital transformation and the rise of immobility**

Mobility has long been regarded as essential to participation, inclusion, and well-being (De Vos et al., 2013; Urry, 2002). However, the expansion of digital technologies and platform economies has profoundly restructured the spatial and temporal organization of everyday life (Rosa et al., 2017; Wajcman & Dodd, 2016; Zheng & Wu, 2022). Increasingly, online platforms substitute for physical movement, enabling individuals to meet daily needs, access services, and engage in economic and social activities without leaving their location. Digital services such as remote work and instant delivery blur the boundaries between domestic, professional, and consumer spaces, allowing for more flexible configurations of time use (Keeble et al., 2020; Wajcman, 2020). Remaining in place can represent a deliberate strategy of time–space management, producing both psychological and economic benefits(Traynor et al., 2022; Wang & He, 2021). This shift reduces reliance on movements across geographic space and fosters new forms of *immobility*. In this context, food delivery services exemplify "spatial substitution," where physical travel is replaced by virtual ordering, transforming immobility from a passive condition into an economic and social capability—what Xiang (2021) termed *immobility capital*. Immobility, in this sense, can be equally functional and productive as mobility.

Research on urban freight logistics has largely focused on the externalities associated with trucks, rail, and both inter- and intra-urban freight transport (Baker et al., 2023; Charters-Gabanek et al., 2025; Cui et al., 2015). Commonly reported negative impacts include traffic congestion, truck-related accidents, greenhouse gas emissions, local air pollution, and urban sprawl (Lezcano et al., 2023; Sanchez-Diaz et al., 2021; Yuan, 2018; E. Zhang et al., 2025). As Yuan and Zhu (2019) noted, the spatial reorganization of warehousing facilities and activities has led to urban sprawl in Wuhan, China and potentially attract more concentrated truck activities, contributing to environmental externalities and disproportionately affecting socially disadvantaged population.

In contrast, far fewer studies have examined the externalities emerging from immobility as facilitated by on-demand delivery services. These services generate impacts that parallel traditional freight logistics but arise from fundamentally different mechanisms. Compared to conventional freight logistics designed for mass production and bulk movement, on-demand delivery services celebrate a model of instant demand. These services rely on digital platforms to algorithmically match, mobilize, and coordinate labor and resources in real time, enabling the rapid transfer of goods directly from retailers to consumers (Huang, 2023; Zheng & Wu, 2022). This hyper-responsive system reconfigures spatial and temporal relations, not only reshaping consumption patterns but also amplifying pressures on labor, infrastructure, and the urban environment (Fan et al., 2017; Huang, 2022; Lezcano et al., 2023; Sanchez-Diaz et al., 2021; Sun, 2019; Van Doorn, 2017).

Historically, the widespread adoption of private vehicles, driven by falling car and fuel prices, has imposed societal costs that far exceed the costs borne by drivers (Banister, 2008; Brown et al., 2009). Similarly, the proliferation of low-cost delivery services is reshaping urban lifestyles, the built environment, and the ways individuals manage space and time (Ma et al., 2024), yet its broader socio-environmental consequences, particularly in relation to the pursuit of speed. When speed is examined, research has largely centered on system efficiency-oriented concerns such as optimizing delivery routes, timing, and pricing (e.g., Ma et al., 2025; Zhou, Liu, et al., 2024). However, strategies for addressing the externalities generated by ever-faster and increasingly voluminous on-demand deliveries remain largely unexplored.

**2.2 Pricing in urban transportation**

Pricing has been promoted widely to correct the externalities of urban transportation(Anas & Lindsey, 2011; Pierce & Shoup, 2013; Taylor, 2006; Wachs, 1981). Road pricing, congestion charges, and vehicle-mile taxation aim to align private costs with social costs by internalizing negative externalities such as congestion, emissions, accidents, and urban sprawl (e.g., Emmerink et al., 1995; Small & Gómez-Ibáñez, 2005; Zhong & Li, 2023). The central rationale is that user fees can both regulate demand and generate revenue to offset social harms. Pricing measures often face political debates over equity and strong public resistance, which complicates their implementation (Levinson, 2010; Taylor & Kalauskas, 2010; Verhoef et al., 1997). However, as Manville and Goldman (2018) noted that the revenues generated from pricing can be redistributed to compensate disadvantaged groups, free roads, by contrast, impose uncompensated social costs on those they harm. More recent studies show that the adoption of congestion pricing in New York has demonstrated tangible benefits, including reduced traffic congestion and improved air quality (Cook et al., 2025; Ghassabian et al., 2024).

Yet, in the case of on-demand delivery services, pricing strategies have primarily reflected production and operational costs—such as routing, timing, and platform fees—rather than broader social costs (Li & Liang, 2022; Tong et al., 2020; Zhou, Zhu, et al., 2024). Moreover, competition in the emerging immobility business encourages commercial platforms to reduce delivery fees, in some Chinese cities to nearly zero. While such price competition may stimulate demand, it also externalizes costs onto society. By masking the true social costs of hyper-fast delivery, platforms create a distorted pricing system that prioritizes market expansion over long-term sustainability.

Putting a price on immobility remains elusive. Typically, pricing strategies follow two approaches: correcting underpriced uses by aligning prices with true costs, or determining the desired level of demand and then setting prices to achieve it (Goodwin, 2001). It's being called:

price and quantity strategies to correcting externalities (Weitzman, 1974). However, delivery pricing faces three difficulties. First, unlike mobility pricing, where charges are imposed directly on travelers, delivery pricing would need to target consumers who demand the service rather than the couriers who perform the movement (Xiang, 2024; Xiang et al., 2023). Delivery workers being treated more as algorithmically managed commodity than labor obscures the externalized costs borne by gig workers, such as precarious wages, unsafe conditions, and lack of protections(Wood et al., 2019). Second, the marginal social costs of delivery would vary across roads, times of day, and cities (Bickel et al., 2006), yet comprehensive estimates for these costs remain unavailable. Third, even with such estimates in place, setting an appropriate price is challenging because the benefits consumers obtain from delivery services are not directly observable; as with changes in quality or taste, delivery fees fail to reflect these variations (Deaton, 1998). The estimated benefits are critical for designing quantity strategies.

Against this backdrop, stated preference approaches provide a way to uncover consumers' perceived value of delivery services and the welfare implications of access and speed. Unlike market prices, which are distorted by subsidies, competition, and platform strategies, stated preference approaches can capture users' own valuation of service quality, waiting time, and reliability. In hedonic terms, these valuations can be viewed as functions of both directly associated goods (e.g., delivery speed, reliability) and complementary systems (e.g., digital connectivity, urban infrastructure). By quantifying these valuations, policymakers can better assess whether delivery services are underpriced relative to their externalities and design interventions.

## 2. Empirical Context and Methodology

### 3.1 Data Collection

For this study, we selected Beijing (China) as a case study. With over 20 million residents, Beijing cannot represent all Chinese cities; however, its jobs–housing spatial mismatch and increasingly auto-oriented development create strong demand for tools that help residents coordinate daily activities. Platform-based food delivery services have emerged to meet this demand. The low delivery fees have further encouraged their widespread adoption across Chinese cities. Figure 1 illustrates that, for the same food ordered at different times, delivery fees range from ¥1.4 to ¥2.4, which is substantially lower than the food price of ¥68 (approximately 2 percent of food price). By analogy to the transportation costs in Glaeser and Kohlhase (2004), one would argue the delivery fees are virtually zero.

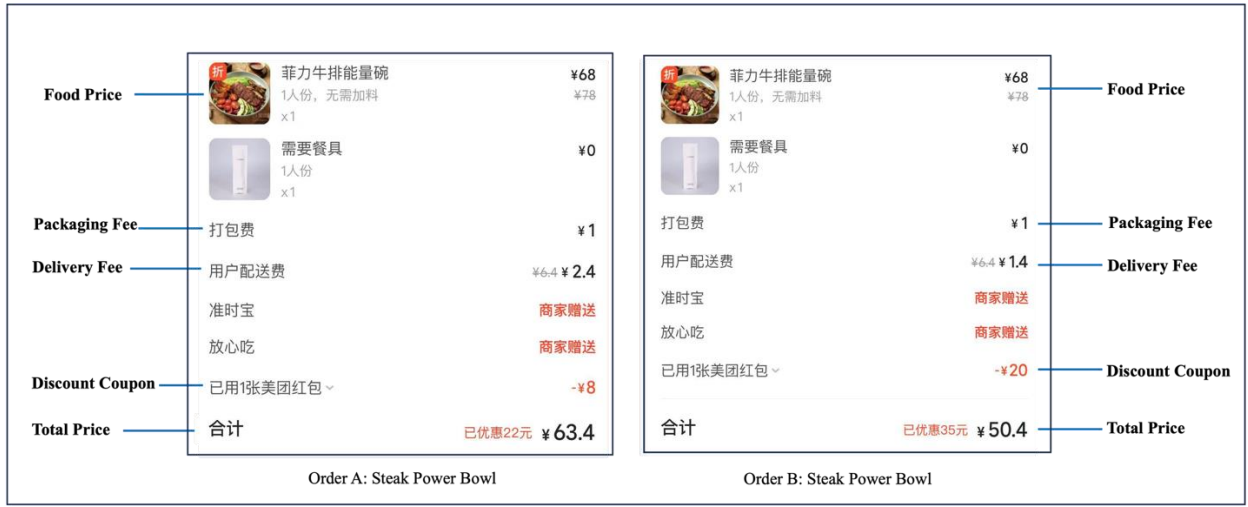

Figure 1. Fees and discounts for ordering food deliveries

Note: Orders of the same food by the authors

Our sampling framework has two screening criteria: (1) respondents must be residents of Beijing, and (2) they must have a consistent history of using food delivery services. Respondents for our online survey were recruited by a professional panel company, Wenjuanxing (wjx.com). The survey was conducted between 5 and 13 November 2024. In total, 598 questionnaires were collected. After applying a series of quality-control procedures—including consistency checks, exclusion of incomplete responses, and a minimum completion time threshold—525 valid questionnaires were retained for subsequent analysis.

Compared with official statistics for Beijing reported in the *China Statistical Yearbook 2024*, the sample is broadly consistent in gender, hukou distribution, marital status and employment type. However, the age profile skews younger, and respondents report higher levels of education (Table 1). This pattern reflects the characteristics of active food delivery users, who are primarily younger and better-educated groups, rather than sampling bias.

**Table 1. Sample Description**

| Variable | | Sample Number | Sample Share | 2024 China Statistical Year Book (Beijing) |
|---|---|---|---|---|
| Gender | Male | 247 | 47.33% | 51.04% |
| | Female | 278 | 52.67% | 48.96% |
| Hukou | Agricultural Hukou | 135 | 25.71% | 12.17% |
| | Urban Hukou | 390 | 74.29% | 87.83% |
| | Local Hukou | 376 | 71.62% | 64.58% |
| | Non-local Hukou | 149 | 28.38% | 35.42% |
| Age | 18-24 | 41 | 7.81% | 4.97% |
| | 25-34 | 286 | 54.48% | 13.33% |
| | 35-44 | 168 | 32.00% | 14.82% |
| | 45-54 | 25 | 4.76% | 15.45% |
| | 55-64 | 2 | 0.38% | 13.86% |
| | Above 65 | 3 | 0.57% | 15.40% |
| Education Level | Primary School | 1 | 0.19% | 10.62% |
| | Middle School | 3 | 0.57% | 20.38% |
| | High School | 12 | 2.29% | 17.54% |
| | Secondary/Technical/Vocational School | 10 | 1.90% | 15.61% |
| | Associate Degree | 30 | 5.71% | |
| | Bachelor's Degree | 401 | 76.19% | 25.82% |
| | Master's Degree or Above | 68 | 12.95% | 8.72% |
| Employment Type | Public Sector | 31 | 5.90% | 32.99% |
| | State-owned Enterprise | 100 | 19.05% | |
| | Private Enterprise | 328 | 62.46% | 54.29% |
| | Foreign Enterprise | 38 | 7.24% | 12.71% |
| | Self-employed | 9 | 1.71% | / |
| | Other | 19 | 3.62% | |
| Income | Below ¥4000 | 20 | 3.81% | |
| | ¥4000 - ¥8000 | 87 | 16.57% | |
| | ¥8000 - ¥12000 | 192 | 36.57% | / |
| | ¥12000 - ¥16000 | 112 | 21.33% | |
| | ¥16000 - ¥20000 | 74 | 14.10% | |
| | Above ¥20000 | 40 | 7.62% | |
| Marital Status | Single | 87 | 16.57% | 20.37% |
| | Married, Living with Spouse | 425 | 80.95% | 71.99% |
|  | Married, Not Living with Spouse | 9 | 1.71% |  |
|  | Divorced/Widowed | 4 | 0.76% | 7.64% |
| Number of Children | 0 | 110 | 20.95% |  |
|  | 1 | 336 | 63.81% |  |
|  | 2 | 65 | 12.38% | / |
|  | 3 | 14 | 2.67% |  |
|  | Above 3 | 0 | 0% |  |
| Number of Elderly | 0 | 370 | 70.48% |  |
|  | 1 | 69 | 13.14% |  |
|  | 2 | 79 | 15.05% | / |
|  | 3 | 7 | 1.33% |  |
|  | Above 3 | 0 | 0% |  |

Note: Population statistics for Beijing was compiled by authors from *China Statistical Yearbook 2024*, *https://www.stats.gov.cn/sj/ndsj/2024/indexch.htm*.

### 3.2 Survey Design

### 3.2.1 Binary Stated Choice Experiment (SBDC): Valuing Service Access

The SBDC experiment presents respondents with a hypothetical situation (Table 2): they had to choose between two options—(a) forgo access to food delivery services for one month in exchange for a specific amount of compensation, or (b) refuse compensation and continue using the service as usual. Since there is no established theoretical guidance on setting compensation levels, we conducted preliminary research and selected a range that is both sufficiently wide and practically reasonable to capture heterogeneous preferences. Each respondent was then randomly assigned four compensation amounts ranging from ¥100 to ¥3,000. Grounded in the framework of perceived value of usage (Moody et al., 2021), the design highlights the extent to which food delivery is embedded in consumers' everyday routines.

**Table 2. Design of the Binary Stated Choice Experiment**

| |
|---|
| **Option A:** Continue to use all food delivery services as usual over the next month. |
| **Option B:** Forgo access to all food delivery services for the next month and receive a cash compensation. *Randomly draw from [100, 500, 1000, 1500, 2000, 2500, 3000]* |

SBDC questions are compatible with economic theory and random utility models. Therefore, we can use a logistic regression to measure the surplus that individual consumers obtain from access to different options and the monetary value that they attach to them.

**3.2.2 Multi-Attribute Stated Choice Experiment (SCE): Valuing Service Attributes**

The SCE experiment was designed to elicit trade-offs between waiting time, time unreliability, and monetary cost (Table 3), with attribute levels informed by a pilot survey and market conditions. A full factorial would yield 351 two-alternative tasks, but we excluded dominated pairs to avoid mechanical choices. From the remaining pool, we generated an efficient fractional design using the *idefix* package in R, ensuring orthogonality and attribute balance. Finally, 16 representative tasks were selected and subjected to manual review and pilot testing to confirm that all scenarios were realistic, consistent, and cognitively manageable.

We incorporated two distinct scenarios: ordering food at work and at home. These settings were chosen because they represent the two primary contexts in which consumers rely on food delivery services, and their decision-making may differ depending on whether meals are ordered during working hours or at home. Each respondent was randomly assigned four choice tasks per scenario. This enables us to find preference structures across different contexts and to estimate individuals' willingness to pay (WTP) for reductions in waiting time, improvements in reliability, and lower costs (Li et al., 2010; Shull et al., 1981).

**Table 3. Attributes and Attribute Levels in Multi-Attribute Stated Choice Experiment**

| Attributes | Levels | | |
|---|---|---|---|
| Waiting time | 30 minutes | 60 minutes | 90 minutes |
| Cost | ¥50 | ¥100 | ¥150 |
| Time Unreliability | ±5 minutes | ±10 minutes | ±15 minutes |

**3.3 Analytic Approach**

We estimate a series of discrete choice models to quantify both the willingness to accept compensation (WTAC) for relinquishing food delivery services and the willingness to pay (WTP) for improvements in service attributes. We begin with two logistic regressions: a baseline specification including only the compensation variable and a random intercept, and an extended model that incorporates socio-demographic characteristics. In both cases, the WTAC is derived from the compensation level at which respondents are equally likely to give up or retain delivery access (i.e., where the probability of acceptance is 0.5). The technical details of this calculation are provided in Appendix A1. This provides a benchmark estimate for the full sample as well as individual-specific thresholds, enabling systematic comparisons across subgroups.

Building on this, we analyze preferences for delivery attributes using choice experiment data. A conditional logit model links respondents' decisions to key service characteristics—waiting time, time reliability, and cost—and allows us to derive marginal WTP estimates in monetary terms (Appendix A2). We also employ a generalized multinomial logit model to capture heterogeneity in preferences (Appendix A3). It allows parameters to vary across individuals, which improves model flexibility and reveals how different groups value time and reliability in food delivery services. We also apply a latent class logit model to capture discrete heterogeneity in preferences (Appendix A4). This approach assumes that respondents cluster into unobserved subgroups with distinct preference structures, allowing us to link differences in valuations of time and reliability to underlying socio-demographic traits.

While individuals' willingness to pay for shorter delivery times captures the private value of food delivery services, it does not automatically reflect the broader social benefits. So we build on the framework proposed by Galvez and Jara-Diaz (Galvez & Jara-Díaz, 1998), which considers both the marginal utility of income and the social weight assigned to different population groups. Under this approach, private WTP can be adjusted to account for income differences and societal priorities, yielding a measure of the social price of time (SPT).

The key insight is that the difference between private valuations and social time prices essentially reflects a choice between equal weighting of all individuals and weighting based on income. The detailed computational steps and formulas are provided in Appendix A5. By incorporating these adjustments, we can translate individual-level time savings into aggregate social welfare gains, providing a more comprehensive assessment of the societal value of immobility enabled by food delivery services.

## 4. Results

### 4.1 Valuation of Access to Food Delivery Services

This section assesses the monetary value consumers place on access to food delivery services, using evidence from the SBDC experiment. Respondents were asked whether they would forgo delivery services for one month in exchange for varying levels of compensation. Their choices provide estimates of WTAC, which we interpret as a measure of the perceived value of access.

The results in Table 4 show a clear positive relationship between compensation levels and respondents' willingness to forgo food delivery services. This highlights the substantial economic and behavioral value consumers place on continued access. In the base model, the median willingness to accept compensation is ¥588 per month, implying that consumers would require at least this amount to give up food delivery for one month.

In a more detailed model, we incorporated variables related to travel behavior, household structure, individual demographics, and patterns of digital consumption. The results reveal that the perceived value of food delivery services is significantly influenced by factors such as income, gender, possession of a driver's license, number of elderly or children in the household, frequency of online shopping and food delivery usage, online shopping, and whether the respondent primarily travels by car.

**Table 4. Results for Valuation of Access to Food Delivery Services in Base Model and Extended Model with Socio-demographic Controls**

|  | Variable | Coefficient | Std. Error |
|---|---|---|---|
| **Base Model** | Compensation (log) | 0.961*** | 0.052 |
|  | Constant | -6.132*** | 0.364 |
| **Extended Model with Socio-demographic Controls** | Compensation (log) | 1.096*** | 0.058 |
|  | Income | 0.185*** | 0.051 |
|  | Age | -0.008 | 0.009 |
|  | Gender | -0.366*** | 0.117 |
|  | Driving License | -0.692*** | 0.252 |
|  | Urban Hukou | 0.026 | 0.143 |

| | | |
|---|---|---|
| Local Hukou | 0.101 | 0.136 |
| Marital Status | -0.071 | 0.189 |
| Education Level | -0.022 | 0.079 |
| Employment Type | 0.060 | 0.069 |
| Number of Children | -0.181* | 0.097 |
| Number of Elderly | -0.240*** | 0.069 |
| Housework Responsibility | -0.183 | 0.130 |
| Commuting Cost | -0.048 | 0.041 |
| Child Pickup Time | 0.004 | 0.057 |
| Online Shopping Frequency | -0.216*** | 0.082 |
| Food Delivery Frequency | -0.398*** | 0.083 |
| Online Shopping Expenditure | -0.001*** | 0.000 |
| Food Delivery Expenditure | -0.000 | 0.000 |
| Car as Primary Transport | -0.493*** | 0.167 |
| Constant | -3.906*** | 0.831 |

Note: $p < 0.1$ (*), $p < 0.05$ (**), $p < 0.01$ (***)

Moreover, each individual possesses unique socio-demographic characteristics. Based on this, we derive a personalized "indifference compensation" value, denoted as $C_i$, by solving for the compensation amount at which the probability of choosing to forgo food delivery services equals 0.5. Rather than yielding a single point estimate, this approach generates a distribution of $C_i$ values across the sample at the indifference threshold. The median value of this distribution is calculated to be ¥589.72, which closely aligns with the ¥588 estimated in the base model. This similarity indicates the robustness of the valuation for food delivery service usage.

**4.2 Valuation of Delivery Service Attributes**

We estimated preferences in both work and home scenarios using Conditional Logit and Generalized Multinomial Logit models to examine how consumers trade off delivery attributes across different contexts (Table 5). In both settings, delivery time, cost, and unreliability exert statistically significant negative effects on consumer choices. WTP estimates from the Conditional Logit models show that consumers are willing to pay around ¥96.6 (work) and ¥99.6

(home) to reduce delivery time by one hour, and ¥4.83 (work) versus 6.68 (home) to improve reliability by one minute (Table 6).

The GMNL results confirm the robustness of these patterns while highlighting heterogeneity in preferences. Consumers are willing to pay approximately ¥92.4 (work) and ¥90.6 (home) per hour to reduce delivery time, but place a much higher value on reliability, with WTP reaching ¥12.95 (work) and ¥19.40 (home) per minute. It's noted that although the direction for the home context aligns with expectations, it does not reach statistical significance and should be treated as suggestive only.

Across both work and home contexts, delivery time, cost, and reliability consistently shape consumer choices. Yet, reliability emerges as relatively more important in the home setting, while time savings play a stronger role in the workplace, underscoring that service predictability is valued most strongly, particularly in the home context. These differences highlight how context influences the way consumers value food delivery services.

**Table 5. Results of Valuation of Valuing Service Attributes in Work and Home Scenarios**

| Scenario | Model | Variable | Coefficient | Std. Error |
|---|---|---|---|---|
| Work Scenario | Conditional Logit | Waiting time | -0.034*** | 0.004 |
| | | Cost | -0.021*** | 0.002 |
| | | Time Unreliability | -0.102*** | 0.022 |
| | Generalized Multinomial Logit | **Mean** | | |
| | | Waiting time | -0.091** | 0.034 |
| | | Cost | -0.059** | 0.022 |
| | | Time Unreliability | -0.760** | 0.328 |
| | | **SD** | | |
| | | Waiting time | -0.043** | 0.020 |
| | | Time Unreliability | -0.659** | 0.263 |
| | | /tau | 1.327*** | 0.254 |
| Home Scenario | Conditional Logit | Waiting time | -0.032*** | 0.004 |
| | | Cost | -0.061*** | 0.002 |
| | | Time Unreliability | -0.217*** | 0.023 |
| | Generalized Multinomial Logit | **Mean** | | |
| | | Waiting time | -0.101 | 0.070 |

|  |  |  |
|---|---|---|
| Cost | -0.067 | 0.042 |
| Time Unreliability | -1.303* | 0.668 |
| **SD** |  |  |
| Waiting time | 0.053* | 0.026 |
| Time Unreliability | -0.837* | 0.434 |
| /tau | 1.604*** | 0.254 |

Note: p < 0.1 (*), p < 0.05 (**), p < 0.01 (***)

**Table 6. Consumers' Willingness to Pay of Waiting Time and Reliability Across Different Scenario**

| Attribute | Regression Model | Work Scenario | Home Scenario |
|---|---|---|---|
| WTP to Waiting Time | Conditional logit | 96.6 ¥/hour | 99.6 ¥/hour |
|  | Generalized Multinomial Logit | 92.4 ¥/hour | 90.6 ¥/hour |
| WTP to Time Reliability | Conditional logit | 4.83 ¥/minute | 6.68 ¥/minute |
|  | Generalized Multinomial Logit | 12.95 ¥/minute | 19.40 ¥/minute |

**4.3 Latent Class Analysis of Preference Heterogeneity**

We further applied a Latent Class Logit Model to examine heterogeneity in consumer preferences by uncovering unobserved subgroups. Models with two to five classes were estimated and compared using fit statistics, including AIC and BIC (details are reported in Appendix 5. Table A1). While additional classes improved fit only marginally, they increased model complexity and reduced interpretability. Therefore, the two-class specification was selected as the most parsimonious and informative solution.

The model identifies two distinct consumer segments: Efficiency Seekers (Class 1, 35.1%) and Reliability Seekers (Class 2, 64.9%). Efficiency Seekers are highly sensitive to delivery time and cost but are relatively indifferent to delivery unreliability. In contrast, Reliability Seekers prioritize price and reliability, demonstrating greater tolerance for longer delivery times. Membership in these segments is influenced by key demographics. Efficiency Seekers tend to be higher-income (≥¥8,000/month), middle-aged (40-60), and possess an agricultural hukou. They also typically spend more per order (over ¥250). Conversely, Reliability Seekers are more often characterized by a non-local hukou and possession of a driver's license—a trait that may indicate less reliance on delivery services, thereby increasing their focus on reliability.

**Table 8. Effects of Socio-Demographic Variables on Latent Class Assignment**

|  | Variable | Coefficient | Std. Error |
|---|---|---|---|
| Choice1 | Waiting time | -0.025*** | 0.005 |
|  | Cost | -0.011*** | 0.003 |
|  | Time Unreliability | 0.009 | 0.028 |
| Choice2 | Waiting time | -0.019 | 0.016 |
|  | Cost | -0.021*** | 0.008 |
|  | Time Unreliability | -0.673*** | 0.013 |
| Class 1 | Income: 8000-16000 | 1.261*** | 0.481 |
|  | Income: above 16000 | 1.157** | 0.534 |
|  | Gender: man | -0.154 | 0.314 |
|  | Age: 40-60 | 0.412* | 0.235 |
|  | Age: above 60 | -50.521 | 31.047 |
|  | Education: College or above | -1.362 | 1.067 |
|  | Hukou: Local Agricultural Hukou | 1.729*** | 0.563 |
|  | Hukou: Non-local Urban Hukou | -1.340*** | 0.484 |
|  | Hukou: Non-local Agricultural Hukou | -1.160** | 0.501 |
|  | Marriage: Married | 0.233 | 0.455 |
|  | Living with Children | 12.609 | 149.918 |
|  | Living with Elderly | -14.481 | 150.042 |
|  | Driver's License | -0.830* | 0.641 |
|  | Housework Responsibility | 0.039 | 0.338 |
|  | Car as Primary Transport | -0.558 | 0.446 |
|  | Commuting Cost > ¥150 | 0.431 | 0.407 |
|  | Child Pickup Time | 0.497* | 0.329 |
|  | Online Shopping Frequency ≥ 3 Times/Week | 0.084 | 0.346 |
|  | Food Delivery Frequency ≥ 3 Times/Week | -0.265 | 0.359 |
|  | Food Delivery Expenditure > ¥250 | 1.356*** | 0.389 |

Note: p < 0.1 (*), p < 0.05 (**), p < 0.01 (***)

## 4.4 Welfare Effects of Delivery Services

Consumers' WTP reflects personal time savings, which the Social Price of Time (SPT) adjusts for income and societal context, with the resulting ΔWs measuring the overall social benefits of reduced waiting times. We choose the results of WTP by the conditional logit model in the work scenario, as a proxy for the subjective value of waiting. The results indicate a WTP of ¥96.6 per hour, chosen because work-time valuations are more consistent and broadly representative. We further estimate the SPT based on WTP, providing a macro-level assessment of the social benefits generated by immobility time in food delivery services.

We estimate the SPT by first assigning each respondent to an income bracket and using the midpoint of that bracket as a representative income value. We then combine these representative income values with individuals' WTP for reduced delivery times, adjusting for differences in the marginal utility of income, to derive the SPT for each income group. This process translates private valuations of time into a societal metric, reflecting both income disparities and the broader social value of time savings. Detailed calculation procedures and underlying formulas are provided in Appendix A5.

**Table 9. Social Price of Time Across Income Groups**

| Income Bracket (¥/month) | Income Value (¥) | SPT |
|:---:|:---:|:---:|
| Under 4000 | 2000 | 19.32 |
| 4000-8000 | 6000 | 57.96 |
| 8000-12000 | 10000 | 96.6 |
| 12000-16000 | 14000 | 135.24 |
| 16000-20000 | 18000 | 173.88 |
| Above 20000 | 22000 | 212.52 |

The results indicate a positive association between SPT and income based on collected sample (Table 9). For higher-income individuals, the greater opportunity cost of time translates into a higher social value per unit saved. In contrast, although lower-income groups may subjectively value their time in similar ways, the economic valuation of their time in policy contexts remains markedly lower. This divergence points to a form of "temporal inequality" in social evaluation and underscores the need for policy frameworks that account more equitably for time across income groups.

Next, we calculate the change in social welfare (ΔWs) by aggregating the SPT across all income groups and weighting it by the size of each group in the sample, which helps assess the broader contribution of time savings enabled by food delivery services. ΔWs reflects the total societal gains from reducing waiting times (Table 10).

The analysis estimates the change in social welfare for our sample, amounting to ¥60,504 per hour, or about ¥1,008 per minute. Middle-income groups account for the largest share of this gain, while the higher marginal utility of time among low-income individuals translates into a smaller aggregate contribution due to their limited population size. These results caution against treating the value of time as uniform across income groups, as doing so risks underestimating the relative benefits for lower-income populations and may introduce bias into resource allocation and policy design.

**Table 10. Social Welfare Change Across Income Groups**

| Income Bracket (¥/month) | Sample Size | Income Value (¥) | Social Welfare Change (¥/hour) |
|---|---|---|---|
| Under 4000 | 20 | 2000 | 386.40 |
| 4000-8000 | 87 | 6000 | 5052.12 |
| 8000-12000 | 192 | 10000 | 18547.20 |
| 12000-16000 | 112 | 14000 | 15123.84 |
| 16000-20000 | 74 | 18000 | 12899.28 |
| Above 20000 | 40 | 22000 | 8494.80 |
| Total | 525 | / | 60,504 |

## 5. Conclusions

The externalities of urban food delivery services have widely been recognized. However, most studies to date have not questioned whether customers pay enough to internalize the social costs. Here, we conduct two experiments to quantify the value of access and use urban food delivery services. The SBDC experiment reveals that consumers place a substantial value on access to the food delivery services (median value ¥588). In the MDC experiment, the estimated willingness to pay for waiting time saving and reliability improvements substantially exceeds typical delivery fees. The value of waiting time is ¥96.6/hour (work scenario) and ¥99.6/hour (home scenario), and the value of reliability is ¥4.83/min (work scenario) and ¥6.68/min (home scenario). These findings echo long-standing dilemmas in transport pricing, where underpricing leads to social inefficiencies.

Despite the stated-preference method's limitations—such as potential divergence from real-world behavior—and the study's focus on a single metropolitan area, the results provide a robust baseline for future assessments. Moreover, tracking changes in consumer valuations over time may prove more policy-relevant than current absolute values.

Most importantly, this study demonstrates the viability of a quantity-based pricing approach for correcting delivery-related externalities. Specifically, we propose pricing delivery speed rather than order volume. Since excessive speed drives many externalities—and volume pricing is both politically and economically infeasible—speed-based pricing offers a targeted, scalable mechanism. By setting price adjustments proportional to consumer surplus, policymakers can reduce negative externalities while preserving net welfare gains.

# Appendix

## A1. Logistic Regression Models for Willingness-to-Accept Compensation (WTAC)

We employ two logistic regression specifications to evaluate the WTAC for access to transportation options across scenarios. The first model is the base model which incorporates only the compensation variable and a random intercept. The second extends the framework by adding socio-demographic controls, capturing how individual and household characteristics shape the likelihood of relinquishing access to food delivery services.

1) Base Model

In the base model, we model the decision to forgo access to a food delivery service option t (choice = 1) in return for a compensation amount C, or to retain access and decline the payment (choice = 0). The binary outcome is expressed as a function of a random intercept and the logarithm of the compensation amount, estimated via logistic regression:

$$\log\left(\frac{p_i}{1-p_i}\right) = \beta_{0,i} + \beta_c \log(C) + \varepsilon_i$$

where $p_i$ denotes the probability that individual $i$ relinquishes service access. The mean of the random intercept $\beta_{0,i}$ and the coefficient $\beta_c$ are estimated using the *pglm* package in R (Croissant, 2020b). The median WTAC is then derived by setting $p_i = 0.5$ and solving for C:

$$\log\left(\frac{0.5}{1-0.5}\right) = \overline{\beta_{0,i}} + \widehat{\beta_c} \log(C)$$

$$\log(1) \text{ or } 0 = \overline{\beta_{0,i}} + \widehat{\beta_c} \log(C)$$

$$-\frac{\overline{\beta_{0,i}}}{\widehat{\beta_c}} = \log(C)$$

$$C = e^{\frac{-\overline{\beta_{0,i}}}{\widehat{\beta_c}}}$$

This yields the compensation level at which consumers are indifferent between giving up and retaining access, providing a point estimate of WTAC for the full sample.

2）Extended Model with Socio-demographic Controls

In the extended model, we incorporate a vector of socio-demographic variables $x_i$ for each respondent. The specification becomes:

$$\log\left(\frac{p_i}{1-p_i}\right) = \beta_{0,i} + \beta_c \log(C) + \boldsymbol{\beta_x} \boldsymbol{x_i} + \varepsilon_i$$

This formulation allows us to estimate an individual-specific compensation threshold. If we when plug in $p_i = 0$ we solve for an individual-specific "indifference compensation" that we call $C_i$. Setting $p_i = 0.5$ and solving for C, we obtain:

$$\log\left(\frac{0.5}{1-0.5}\right) = \widehat{\beta_{0,i}} + \widehat{\beta_c} \log(C) + \widehat{\boldsymbol{\beta_x}} \boldsymbol{x_i}$$

$$C_i = e^{-\frac{\widehat{\beta_{0,i}} + \boldsymbol{\beta_x} \boldsymbol{x_i}}{\widehat{\beta_c}}}$$

Rather than a single point estimate, the model yields a distribution of indifference values $C_i$ across respondents. The median or mean of this distribution can then be used as a sample-level summary, analogous to the baseline WTAC but adjusted for socio-demographic heterogeneity. This approach also enables comparisons across subsamples, highlighting systematic differences in compensation thresholds.

## A2. Conditional Logit Model for Willingness to Pay (WTP)

To analyze preferences for delivery service attributes, we first apply a **conditional logit model** to the multi-attribute choice data derived from the experiment. In this model, the dependent variable represents the choice made by respondent *i* among a set of alternatives *J*. Each alternative *j* is characterized by a vector of service attributes, denoted as $X_{ij}$, which includes waiting time, time reliability, and cost. The probability that individual *i* chooses alternative *j* is given by:

$$\Pr(y_i = j | X_i) = \frac{\exp(\beta_j \cdot X_{ij})}{\sum_{k \in J_i} \exp(\beta_k \cdot X_{ik})}$$

After estimating the marginal utility coefficients, we derived marginal willingness to pay (WTP) by dividing each attribute coefficient by the cost coefficient. This provides a monetary measure of the value respondents place on shorter waiting times and more reliable delivery services.

**A3. Generalized Multinomial Logit Model for Willingness to Pay (WTP)**

We employ a generalized multinomial logit model to capture heterogeneity in preference intensity and decision uncertainty, extending beyond the conditional logit by allowing parameters to vary across individuals. In this model, the choice made by respondent $i$ is denoted as $y_i$, and the utility of alternative $j$ is specified as a function of observed attributes and random components. The utility function includes both fixed-effect variables (cost) and scenario-specific attributes (waiting time and time reliability), represented by $X_i$ and $Z_i$ respectively. Individual-level identifiers are captured through $W_i$, while $\beta_j$ and $\theta_j$ denote fixed and context-dependent coefficients. The random term $\mu_{ij}$ reflects unobserved preference variation across individuals.

$$P(y_i = j | X_i, Z_i, W_i, \theta, \beta) = \frac{\exp(W_i^T \beta_j + Z_i^T \theta_j + \mu_{ij})}{\sum_{k \epsilon J} \exp(W_i^T \beta_k + Z_i^T \theta_k + \mu_{ik})}$$

By incorporating individual-level randomness and adjusting for scale, the GMNL model improves flexibility and predictive accuracy. The estimates further allow us to derive marginal WTP for key attributes.

**A4. Latent Class Model for Preference Heterogeneity**

We employ the GMNL model to capture continuous heterogeneity in preferences at the individual level. We further apply a latent class logit model, which assumes respondents fall into unobserved subgroups with distinct preference structures, thereby providing clearer subgroup-specific insights.

$$\Pr(y_i = j | X_i, \theta_k) = \frac{\exp(\beta_j \cdot X_{ij})}{\sum_{k \epsilon J_i} \exp(\beta_k \cdot X_{ik})}$$

The probability that individual *i* chooses alternative *j* is modeled as a class-weighted average of class-specific choice probabilities. Within each class *k*, utility is determined by a vector of attribute levels $X_i$ and associated class-specific coefficients $\beta_{jk}$. The likelihood of class membership is modeled using socio-demographic variables $Z_i$, with $\gamma_k$ denoting the corresponding parameter vector.

$$P(k|Z_i) = \frac{\exp(\gamma_k \cdot Z_i)}{\sum_{k' \in K} \exp(\gamma_{k'} \cdot Z_i)}$$

$$\Pr(y_i = j|X_i, Z_i) = \sum_{k \in K} p(k|Z_i) \cdot \Pr(y_i = j|X_i, \theta_k)$$

By uncovering systematic preference differences across latent classes, the model allows us to identify distinct consumer groups and link their valuations of immobility-related attributes to underlying social and economic characteristics.

### A5. Welfare Analysis

While individuals' WTP for shorter waiting times captures the private value of immobility, it does not necessarily reflect its broader social benefits. In policy evaluation, distinguishing personal utility from collective welfare is essential. When time saved can be reallocated to productive activities, it acquires a social value—commonly referred to as the Social Price of Time (SPT)—often proxied by wage rates as an indicator of marginal labor productivity.

We build on the framework proposed by Galvez and Jara-Diaz (1998) to estimate the change in social welfare (ΔWs) associated with the valuation of waiting time savings. Specifically, we calculate welfare improvements for each population subgroup *q*, incorporating the group's average WTP for delivery time reductions (SVTT$_s$), the social priority or policy relevance of the group ($\Omega_s$), and the marginal utility of income ($\lambda_s$).

$$\Delta W_s = \sum_q \Omega_q \lambda_q SVTT_q \Delta t_q$$

In this formulation, $SVTT_q$ reflects private WTP, $\Omega_q$ represents the relative social priority or weight given to group $q$, and $\lambda_q$ denotes the marginal utility of income for that group. If $\Omega_q$ is assumed to be inversely related to income (i.e., higher incomes receive lower social weight), then private WTP can directly serve as the SPT. This highlights that using WTP as a basis for social valuation implicitly assumes a regressive distribution of welfare weights.

Alternatively, under the "one person, one vote" principle ($\Omega_q = 1$), $\Delta W_s$ can be expressed in monetary terms by applying a common social monetary utility factor ($\lambda_s$). In this case, the social price of time for group $q$ becomes:

$$SPT_q = \frac{\lambda_s}{\lambda_q} SVTT_q$$

Further simplification shows that when the marginal utility of travel time and the marginal utility of cost are expressed as a ratio equal to SVTT, the social price of time reduces to:

$$SPT_q = \frac{\lambda_s}{|\frac{\partial V_i}{\partial t_i}|_q}$$

The essential insight here is that the difference between private WTP and SPT reflects a normative choice between equal weighting of individuals and weighting based on income. Calculating $\Delta W_s$ for immobility time savings thus allows us to evaluate the societal value of food delivery services beyond private preferences, and to assess the distributive implications of treating time as either a productive resource or a source of personal well-being.

**Table A1 Latent Class Model Fit Statistics**

| Classes | LLF | Nparam | BIC | AIC |
| --- | --- | --- | --- | --- |
| 2 | -777.6628 | 7 | 1599.169 | 1569.326 |
| 3 | -769.1633 | 11 | 1607.224 | 1560.327 |
| 4 | -763.2635 | 15 | 1620.478 | 1556.527 |
| 5 | -760.5145 | 19 | 1640.033 | 1559.029 |